\documentclass{elsart}

\usepackage{graphicx}

\usepackage{amssymb}

\begin{document}
\begin{frontmatter}
%%Juanjo
%\begin{flushright}
%\parbox[b]{0.3\textwidth}
%{\raggedright Version 3.0}
%\end{flushright}
\title{The ANTARES Optical Beacon System}
\renewcommand{\thefootnote}{\fnsymbol{footnote}}
\begin{center}
\author[CPPM]{M.~Ageron},
\author[IFIC]{J.A.~Aguilar\corauthref{cor1}},
\ead{J.A.Aguilar@ific.uv.es}
\corauth[cor1]{Corresponding author}
\author[Mulhouse]{A.~Albert},
\author[Rome]{F.~Ameli},
\author[Genova]{M.~Anghinolfi},
\author[Erlangen]{G.~Anton},
\author[Saclay]{S.~Anvar},
\author[Saclay]{F.~Ardellier-Desages},
\author[CPPM]{E.~Aslanides},
\author[CPPM]{J-J.~Aubert},
\author[Erlangen]{R.~Auer},
\author[Bari]{E.~Barbarito},
% ...

\newpage
\nopagebreak[3]
\address[APC]{APC -- AstroParticule et Cosmologie, 10, rue Alice Domon et L\'eonie Duquet 75205 Paris Cedex 13, France}
\vspace*{-0.20\baselineskip}
\nopagebreak[3]
\address[Bari]{Dipartimento Interateneo di Fisica e Sezione INFN, Via E. Orabona 4, 70126 Bari, Italy}
\vspace*{-0.20\baselineskip}
\nopagebreak[3]
\address[Bologna]{Dipartimento di Fisica dell'Universit\`a e Sezione INFN, Viale Berti Pichat 6/2, 40127 Bologna, Italy}
\vspace*{-0.20\baselineskip}
\nopagebreak[3]
\address[COM]{COM -- Centre d'Oc\'eanologie de Marseille, CNRS/INSU et Universit\'e de la M\'editerran\'ee, 163 Avenue de Luminy, Case 901, 13288 Marseille Cedex 9, France}
\vspace*{-0.20\baselineskip}
\nopagebreak[3]
\address[CPPM]{CPPM -- Centre de Physique des Particules de Marseille, CNRS/IN2P3 et Universit\'e de la M\'editerran\'ee, 163 Avenue de Luminy, Case 902, 13288 Marseille Cedex 9, France}
\vspace*{-0.20\baselineskip}
\nopagebreak[3]
\address[Catania]{Dipartimento di Fisica ed Astronomia dell'Universit\`a e Sezione INFN, Viale Andrea Doria 6, 95125 Catania, Italy}
\vspace*{-0.20\baselineskip}
\nopagebreak[3]
\address[Erlangen]{Friedrich-Alexander-Universit\"at Erlangen-N\"urnberg, Physikalisches Institut, Erwin-Rommel-Str.\ 1, D-91058 Erlangen, Germany}
\vspace*{-0.20\baselineskip}
\nopagebreak[3]
\address[GEOAZUR]{G\'eoSciences Azur, CNRS/INSU, IRD, Universit\'e de Nice Sophia-Antipolis, Universit\'e Pierre et Marie Curie -- Observatoire Oc\'eanologique de Villefranche, BP48, 2 quai de la Darse, 06235 Villefranche-sur-Mer Cedex, France}
\vspace*{-0.20\baselineskip}
\nopagebreak[3]
\address[Genova]{Dipartimento di Fisica dell'Universit\`a e Sezione INFN, Via Dodecaneso 33, 16146 Genova, Italy}
\vspace*{-0.20\baselineskip}
\nopagebreak[3]
\address[IFIC]{IFIC -- Instituto de F\'{\i}sica Corpuscular, Edificios de Investigaci\'on de Paterna, CSIC -- Universitat de Val\`encia, Apdo. de Correos 22085, 46071 Valencia, Spain}
\vspace*{-0.20\baselineskip}
\nopagebreak[3]
\address[IFREMER/Brest]{IFREMER -- Centre de Brest, BP 70, 29280 Plouzan\'e, France}
\vspace*{-0.20\baselineskip}
\nopagebreak[3]
\address[IFREMER/Toulon]{IFREMER -- Centre de Toulon/La Seyne Sur Mer, Port Br\'egaillon, Chemin Jean-Marie Fritz, 83500, La Seyne sur Mer, France}
\vspace*{-0.20\baselineskip}
\nopagebreak[3]
\address[INAF]{INAF-IASF, via P. Gobetti 101, 40129 Bologna, Italy}
\vspace*{-0.20\baselineskip}
\nopagebreak[3]
\address[IPHC]{IPHC -- Institut Pluridisciplinaire Hubert Curien, Universit\'e Louis Pasteur (Strasbourg 1) et IN2P3/CNRS, 23 rue du Loess, BP 28, 67037 Strasbourg Cedex 2, France}
\vspace*{-0.20\baselineskip}
\nopagebreak[3]
\address[ISS]{Institute for Space Sciences, R-77125 Bucharest - M\u{a}gurele, Romania.}
\vspace*{-0.20\baselineskip}
\nopagebreak[3]
\address[ITEP]{ITEP -- Institute for Theoretical and Experimental Physics, B.~Cheremushkinskaya 25, 117259 Moscow, Russia}
\vspace*{-0.20\baselineskip}
\nopagebreak[3]
\address[KVI]{Kernfysisch Versneller Instituut (KVI), University of Groningen, Zernikelaan 25, 9747 AA Groningen, The Netherlands}
\vspace*{-0.20\baselineskip}
\nopagebreak[3]
\address[LAM]{LAM -- Laboratoire d'Astrophysique de Marseille, CNRS/INSU et Universit\'e de Provence, Traverse du Siphon -- Les Trois Lucs, BP 8, 13012 Marseille Cedex 12, France}
\vspace*{-0.20\baselineskip}
\nopagebreak[3]
\address[LNS]{INFN -- Labaratori Nazionali del Sud (LNS), Via S. Sofia 44, 95123 Catania, Italy}
\vspace*{-0.20\baselineskip}
\nopagebreak[3]
\address[Leeds]{School of Physics \& Astronomy, University of Leeds LS2 9JT, UK}
\vspace*{-0.20\baselineskip}
\nopagebreak[3]
\address[Mulhouse]{GRPHE -- Groupe de Recherche en Physique des Hautes Energies, Universit\'e de Haute Alsace, 61 Rue Albert Camus, 68093 Mulhouse Cedex, France}
\vspace*{-0.20\baselineskip}
\nopagebreak[3]
\address[NIKHEF]{Nationaal Instituut voor Kernfysica en Hoge-Energiefysica (NIKHEF), Kruislaan 409, 1098 SJ Amsterdam, The Netherlands}
\vspace*{-0.20\baselineskip}
\nopagebreak[3]
\address[Pisa]{Dipartimento di Fisica dell'Universit\`a e Sezione INFN, Largo B.~Pontecorvo 3, 56127 Pisa, Italy}
\vspace*{-0.20\baselineskip}
\nopagebreak[3]
\address[Rome]{Dipartimento di Fisica dell'Universit\`a "La Sapienza" e Sezione INFN, P.le Aldo Moro 2, 00185 Roma, Italy}
\vspace*{-0.20\baselineskip}
\nopagebreak[3]
\address[Saclay]{DSM/Dapnia -- Direction des Sciences de la  Mati\`ere, laboratoire de recherche sur les lois fondamentales de l'Univers, CEA Saclay, 91191 Gif-sur-Yvette Cedex, France}
\vspace*{-0.20\baselineskip}
\nopagebreak[3]
\address[Sheffield]{Dept.\ of Physics and Astronomy, University of Sheffield, Sheffield S3 7RH, UK}
\vspace*{-0.20\baselineskip}
\nopagebreak[3]
\address[UU]{Universiteit Utrecht, Faculteit Betawetenschappen, Princetonplein 5, 3584 CC Utrecht, The Netherlands}
\vspace*{-0.20\baselineskip}
\nopagebreak[3]
\address[UvA]{Universiteit van Amsterdam, Instituut voor Hoge-Energiefysica, Kruislaan 409, 1098 SJ Amsterdam, The Netherlands}
\vspace*{-0.20\baselineskip}
\end{center}

%\author{The ANTARES collaboration (List of authors)}
%~\\[1cm]
%\author{\corauthref{cor1} J.A. Aguilar}
%\ead{J.A.Aguilar@ific.uv.es}
%\ead[url]{http://ific.uv.es/~aguilars}


%\title{}


%\author{}

%\address{}

\begin{abstract}
% Text of abstract

 ANTARES is a neutrino telescope being deployed in the Mediterranean Sea. It
consists of a three dimensional array of photomultiplier tubes that can
detect the Cherenkov light induced by charged particles produced in the
interactions of neutrinos with the surrounding medium. High angular
resolution can be achieved, in particular when a muon is produced, provided
that the Cherenkov photons are detected with sufficient timing
precision. Considerations of the intrinsic time uncertainties stemming from
the transit time spread in the photomultiplier tubes and the mechanism of
transmission of light in sea water lead to the conclusion that a relative
time accuracy of the order of 0.5 ns is desirable. Accordingly, different
time calibration systems have been developed for the ANTARES telescope. In
this article, a system based on Optical Beacons, a set of external and
well-controlled pulsed light sources located throughout the detector, is
described. This calibration system takes into account the optical properties
of sea water, which is used as the detection volume of the ANTARES
telescope. The design, tests, construction and first results of the two types
of beacons, LED and laser-based, are presented.

\end{abstract}

\end{frontmatter}

% main text
\section{Introduction}
\label{chap: int}

The ANTARES telescope is an underwater neutrino detector being deployed in
the Mediterranean Sea at a depth of 2500~m offshore from Toulon
(France)~\cite{ANTARES}. This detector, which is a first step toward a
km$^3$-scale undersea neutrino telescope, will consist of twelve lines with a
sensitive area for high energy muons of more than ~0.05 km$^2$ for E$_\mu >$
100 TeV. The construction of the ANTARES neutrino telescope started with the
installation of the first lines in 2006, and the detector is scheduled to be
completed by the end of 2007. In addition, a special instrumentation line,
the MILOM~\cite{Milom}, is in operation since Spring 2005. The detection
principle and the main detector components are briefly described in
section~\ref{chap: det}. The precision required in the time determination
and the different time calibration systems are reviewed in section~\ref{chap:
timing}.  The concept of an Optical Beacon system is briefly introduced and
the actual solution adopted by ANTARES is reviewed in section~\ref{chap:
OBsystem}. Detailed descriptions of the LED and Laser Beacon calibration
systems are given in section~\ref{chap: LEDsystem} and
\ref{chap: LASERsystem}, respectively. Some results from the first data taken
with the lines in operation in 2006 are given in section~\ref{chap: results}.
Finally, section~\ref{chap: summary} presents the summary and conclusions.

\section{The ANTARES Neutrino Telescope}
\label{chap: det}
 The ANTARES neutrino telescope uses sea water as the detection medium to
look for extra-terrestrial neutrinos. Most of these neutrinos cross right
through the Earth without interacting. A small fraction of the incoming
neutrino flux, however, interacts with the nucleons that make up the
matter surrounding the detector. In a charged current interaction a high
energy muon neutrino produces a muon which induces Cherenkov light when
crossing a suitable optical medium such as ice or water. Other signatures can
also be detected.

 In order to detect and reconstruct the wavefront of the Cherenkov light,
ANTARES is equipped with 900 Optical Modules (OMs). The OM, the basic optical
unit of ANTARES, consists of a photomultiplier tube (PMT) housed in a
water-pressure resistant glass sphere~\cite{OM}. An exhaustive study of PMTs
was carried out during the R\&D phase which led to the selection of the
14-stage, 10'' Hamamatsu R7081-20 model~\cite{PMT}. Together with the PMT
there is an internal LED for calibration purposes inside the OM. Each group
of three OMs constitutes a storey. All the electronics for one storey are
housed in a pressure resistant titanium container making up the so-called
Local Control Module (LCM). Every OM is read out by an electronics board
housed in the LCM carrying a pair of Analogue Ring Samplers (ARS), the ASIC
chip used for signal processing and digitisation~\cite{ARS}. The ARS provides
the time and amplitude of the signal, both of which are essential to
reconstruct the muon track direction and estimate its energy.

\section{The ANTARES time calibration systems}
\label{chap: timing}
 ANTARES is expected to achieve very good angular resolution ($<0.3^{\circ}$
for muon events above 10~TeV). This pointing accuracy is closely related to
the precision in the determination of the arrival time of the Cherenkov
photons at the PMTs. The relative time resolution between OMs is, therefore,
of utmost importance. It is limited by the transit time spread (TTS) of the
signal in the PMTs ($\sigma \sim 1.3$~ns) and by the scattering and
chromatic dispersion of light in sea water ($\sigma \sim 1.5$~ns for a light
propagation of 40 m)~\cite{SEA,Alain}. The electronics of the ANTARES
detector is designed in order to contribute less than 0.5~ns to the overall
time resolution. Therefore, the time calibration should aim at a precision
below the nanosecond level. To this end, several complementary time calibration systems are implemented in
the ANTARES detector in order to measure and monitor the relative times
between different components of the detector due to, e.g. cable lengths and
electronics delays. These time calibrations are performed by the following systems:

\begin{enumerate}
\item The {\bf internal clock} calibration system. A very precise time
  reference clock distribution system has been implemented in the ANTARES
  detector. It consists of a 20 MHz clock generator on shore, a clock
  distribution system and a clock signal transceiver board placed in each LCM. A
  common clock signal is provided to the ARSs. Synchronised data commands can
  be superimposed on the clock signal, in particular {\it start} and {\it
  stop} commands, which together with a high precision Time to Digital
  Converter (TDC) make up the essential components of the system.  This
  system also includes an echo-based time calibration whereby each LCM clock
  electronics board is able to send back a return signal through the same optical
  path as the outgoing clock signals. This system enables the time
  offsets between all LCM clock boards to be measured
  by recording the propagation delays of
  the return signals of each storey with respect to the original clock signal
  emission time. Measurements in real conditions show a resolution of
  $\sim 0.1$~ns, well within the specifications. The system also
  includes the synchronisation with respect to Universal Time, by
  assigning the GPS timestamp to the data, with a
  precision of about 100~$\mu$s, much better than the required
  precision of $\sim1$~ms. 
   The clock signals are distributed across all detector components from the
   shore up to the clock boards. The remaining path between these
   boards and the PMT photocathodes however requires a different timing
   calibration mechanism.
\item The {\bf internal Optical Module LEDs}. Inside each Optical Module
  there is a blue LED attached to the back of the PMT capable of
  illuminating the photocathode. The LED is an HLMP-CB15 from Agilent
  whose light intensity is peaked at around 470 nm with a FWHM of 15
  nm. These LEDs are used to measure the relative variation of the PMT
  transit time and dedicated runs of this LED calibration system are
  customarily taken~\cite{Milom}. This system is used to calibrate the path
  travelled by the signal starting at the PMT photocathode up to the read-out
  electronics. The effect of the transmission of the light in water is,
  however, not addressed by this calibration method.
\item The {\bf Optical Beacons}. 
   This system allows the relative time calibration of different OMs to be
  determined by means of independent and well-controlled pulsed light
  sources. This system also makes possible to monitor the influence of the
  water on the light propagation. These Optical
  Beacons are the subject of this paper and will be described in detail in
  the following sections.
\item Several thousands of down-going {\bf muon tracks} will be detected per day. The hit time residuals of the
  reconstructed muon tracks can be used to monitor the time offsets of the
  Optical Modules. This methodology will enable an overall space-time
  alignment and calibration cross-checks.
\end{enumerate}

 Prior to the deployment of the lines, all line elements are verified as
functioning correctly in a dedicated dark setup where a time calibration is
carried out after the integration of each sector of the line (a sector is one
fifth of a line). An optical signal is sent to each OM of every storey. The signal is
provided by a Nd-YAG solid state laser that emits intense, short duration
light pulses. The light pulse is attenuated before being sent to the OMs.
The light is guided through an optical fibre to a 1-to-16 optical
splitter. Each of the outgoing fibres is connected to one of the 15 OMs of
the sector. The 16$^{\rm{th}}$ signal is sent to a control module and is
used as a time reference. The resulting information from timing
calibration in the dark setup is used as the reference for the validation of
the in situ timing calibrations. This system is also used to determine
the time calibration of the Optical Beacons.

The time calibration depends on the actual location of the OMs which is
affected by the slow movements of the lines due to underwater currents. An
acoustic positioning system together with a set of compasses and tiltmeters
located along the line, provides the OM position with an accuracy of 10-20~cm
which, in addition to the time calibration, is sufficient for the muon track
reconstruction~\cite{Milom}.

\section{The ANTARES Optical Beacon system}
\label{chap: OBsystem}
 The Optical Beacon system consists of a series of pulsed light sources
distributed throughout the detector. An LED Beacon is composed of several
LEDs, pulsed by dedicated electronic circuits. Those beacons are located,
more or less, uniformly along every detector line so that their light can illuminate all storeys
on the neighbouring lines.
The Laser Beacons use a solid state pulsed
laser whose light is spread out by a diffuser. Laser Beacons are located at the
bottom of a few lines in the so-called Bottom String
Socket. The Laser Beacons illuminate mainly the bottom part of the lines and are located in a stationary position.  The system of Optical
Beacons provides a number of well controlled, pulsed light sources that act
as a reference for time calibration of the detector. The system is able to
closely monitor all the detector components and the sea water. It allows the
monitoring of the absorption and scattering lengths of the sea water.

\section{The LED Beacons}
\label{chap: LEDsystem}
\begin{figure}[htpb]
\begin{center}
\begin{tabular}{c}
\hspace{0 cm}
\includegraphics[width=0.7\linewidth, angle=0]{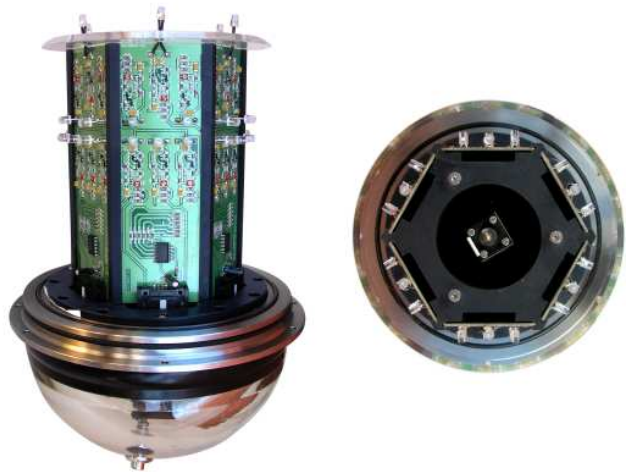}
\end{tabular}
\vspace{-0.1 cm}
\caption{An LED Beacon as viewed from sideways (left) and from the top
(right).  Six faces each containing six LEDs 
are arranged on an hexagonal cylinder. The internal
photomultiplier tube is mounted at the centre of the lightguide.}
\label{fig: LEDBeacon}
\end{center}
\end{figure}

An LED Beacon contains 36 individual LEDs arranged in groups of
six, on six vertical boards ({\it faces}) which are placed side by side
forming an hexagonal cylinder (figure~\ref{fig: LEDBeacon}). On each face,
one LED points upwards ({\it top LED}), and the other five LEDs point radially
outwards. One of the LEDs that points horizontally is located in the middle
of the face ({\it central LED}) and the remaining four surround it.

%%%%%%%%%%%%%%%%%%%%%%%%%
%%%%%%%%%%%%%%%%%%%%%%%%%%%
~\\

 The faces are mechanically fixed to a hollow nylon structure which
internally houses a small Hamamatsu H6780-03 photomultiplier tube. This
PMT, with a photocathode of 8~mm diameter, has a risetime
of 0.8~ns and a transit time of 5.4~ns and is used to provide the precise
time of emission of the light flash independently of the triggering signal.
A flat acrylic disc that acts as a lightguide is fixed to the upper part of
the nylon mounting to increase the collection of light.  Following Fields and
Janowski~\cite{fields} a conical depression was machined in the centre of the
light collecting disc, to direct light into the photomultiplier tube. The edges of
the disc were also bevelled at 45$^{\circ}$ to improve light collection from
the horizontal LEDs.

The lower part of the LED Beacon houses the electronic boards that provide
the required operating voltages and enable the actual LED flashing according
to externally supplied slow control commands.  Each of the six faces can be
flashed independently or in combination of different faces. Within a face the
top, central and the group of four LEDs can be triggered independently or in
combination. This layout allows a distribution close to uniform in the
azimuth angle when all the LEDs are flashing. The top LEDs allow the
calibration of the OMs in those storeys on the same line above the
beacon. The amount of light can be further controlled changing the number of
LEDs flashing at a given time while the intensity of the LEDs can also be
varied as explained in the following subsection.  The possibility of flashing
certain faces enables the monitoring of the uncertainty in the time
calibration arising from the non-uniformity and anisotropy of the LED Beacons.

\subsection{The pulser system}
\label{chap: LBpulser}

\begin{figure}[htpb]
\begin{center}
\begin{tabular}{c}
\hspace{0 cm} \includegraphics[width=0.7\linewidth, angle=0]{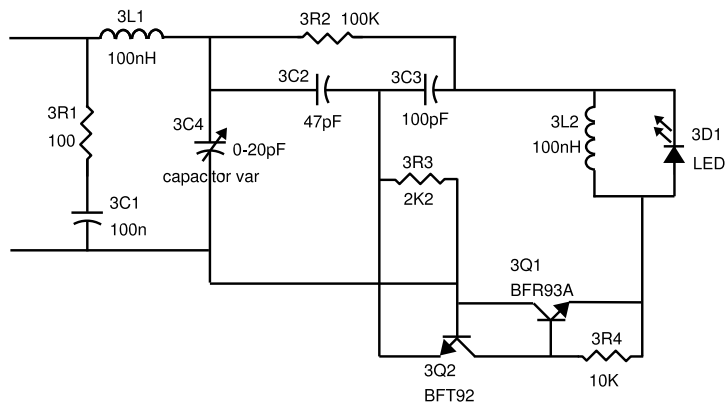}
\end{tabular}
\vspace{-0.1 cm}
\caption{Diagram of the LED pulser circuit. The variable capacitator (see
text) is indicated by label 3C4.}
\label{fig: diagram}
\end{center}
\end{figure}

The pulser circuit is based on an original design from Kapustinsky et
al.~\cite{PULSER} that has been modified for ANTARES to include in particular
a variable capacitor that enables the synchronisation of the pulses produced
by several different circuits (see figure~\ref{fig: diagram}).  The trigger
is provided by a 1.5~V negative square pulse of a duration of around 150~ns
superimposed on a negative DC bias that can be varied from 0 to 24~V. The DC
component charges the capacitor and the rising edge of the differentiated
1.5~V pulse switches on the pair of transistors, triggering the fast
discharge of the 100~pF capacitor through the low impedance path that
includes the LED. The parallel inductor develops charge in opposition to the
discharging capacitor further reducing its time constant. The level of the DC
voltage determines the amount of current through the LED and thus the
intensity of the emitted pulse.  The layout of the printed circuit board has
been designed so as to enable the inclusion of six pulsers on the same face
without interference of the distributed triggering signals in the nearby
pulser circuits whilst minimizing the difference in the times of arrival of
the trigger signal to the different pulsers.  Synchronisation of the signals
from different LEDs on the same or on different faces is possible by
adjusting the variable capacitor in each pulser, this procedure is explained
in subsection~\ref{chap: ASSEMBLY}. When deployed underwater the LED Beacons
are operated with a typical trigger frequency of a few Hz. This frequency can
be increased to rates up to 1~kHz.

%%%%%%%%%%%%%%%%%%%%%%%%%
%%%%%%%%%%%%%%%%%%%%%%%%%%%

\subsection{The LEDs}
\label{chap: LED}
Different types of LEDs were tested by the ANTARES collaboration in terms of
amplitude and risetime of the emitted light pulses.  The selected LED was the
Agilent HLMP-CB15-RSC00 model\footnote{Agilent Technologies,
Inc. Headquarters 395 Page Mill Rd. Palo Alto, CA 94306 United States.}.
This LED has a peak wavelength of 472~nm with a spectral half-width of 35~nm
according to the specification sheet.  The risetime of the LED pulses has
been measured and found to be between 1.9 and 2.2~ns.  The LEDs were classified in
batches according to their risetime and the fastest (1.9-2.0~ns) are used for
the top LED locations where the uncertainty in the calibration is dominated
by the rise-time and not by the light propagation effects. 

To increase the angular occupancy of the light emitted by the LEDs, which was 
originally restricted to $15^\circ$, the caps of
the LEDs were machined off. The angular distribution of the emitted light
for several depths of cut was measured. 
The cut lengths tested ranged from 1.5 to 3.5~mm. 
A cut at 3~mm was selected, which provides an emission flat within 
$\pm 10 \%$ for angles up to $35 ^\circ$ and 
within a factor two up to $55 ^\circ$.

\begin{figure}[htpb]
\begin{center}
\begin{tabular}{c c}
\hspace{0 cm}
\includegraphics[width=0.5\linewidth, angle=0]{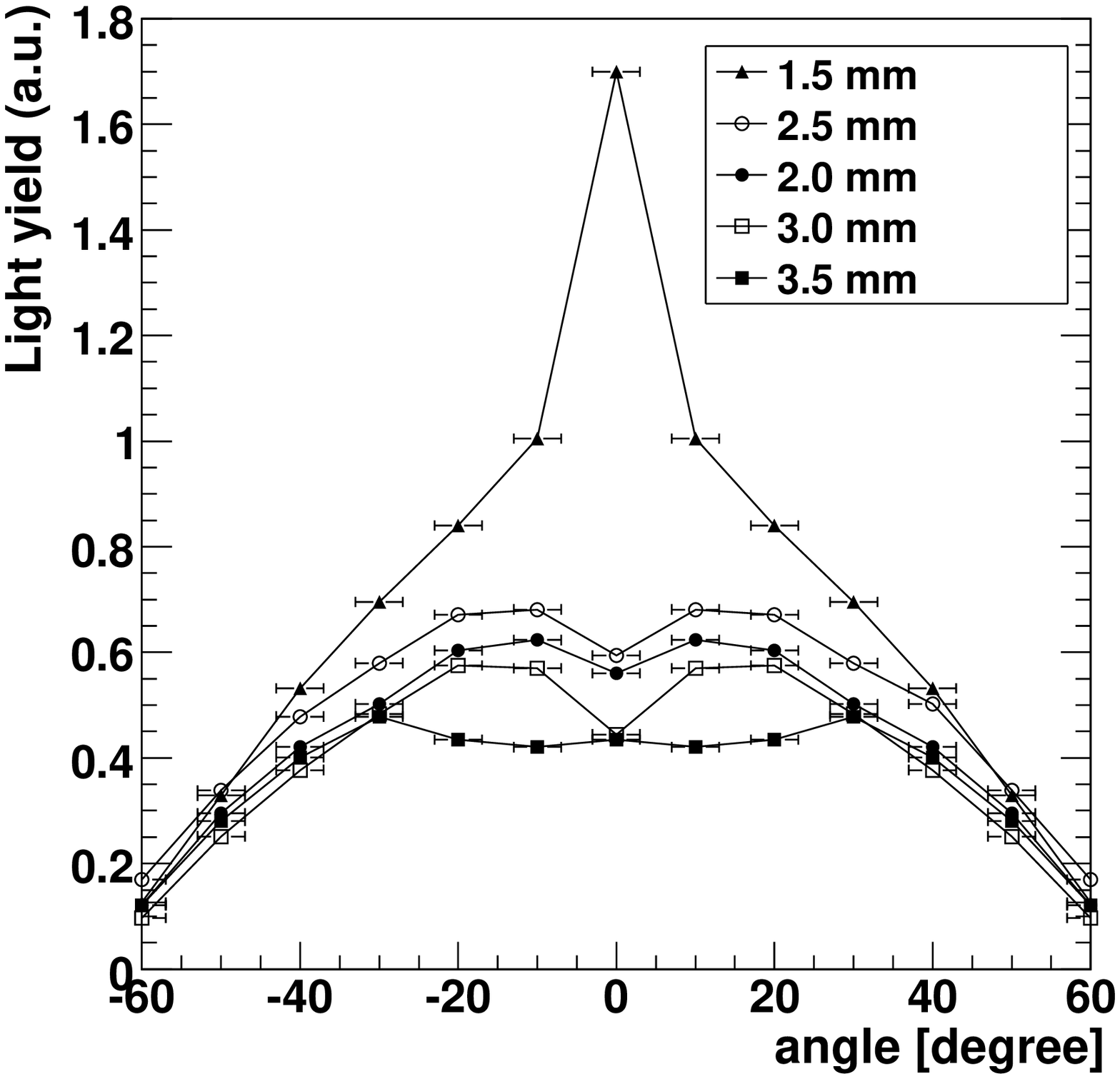}
\includegraphics[width=0.5\linewidth, angle=0]{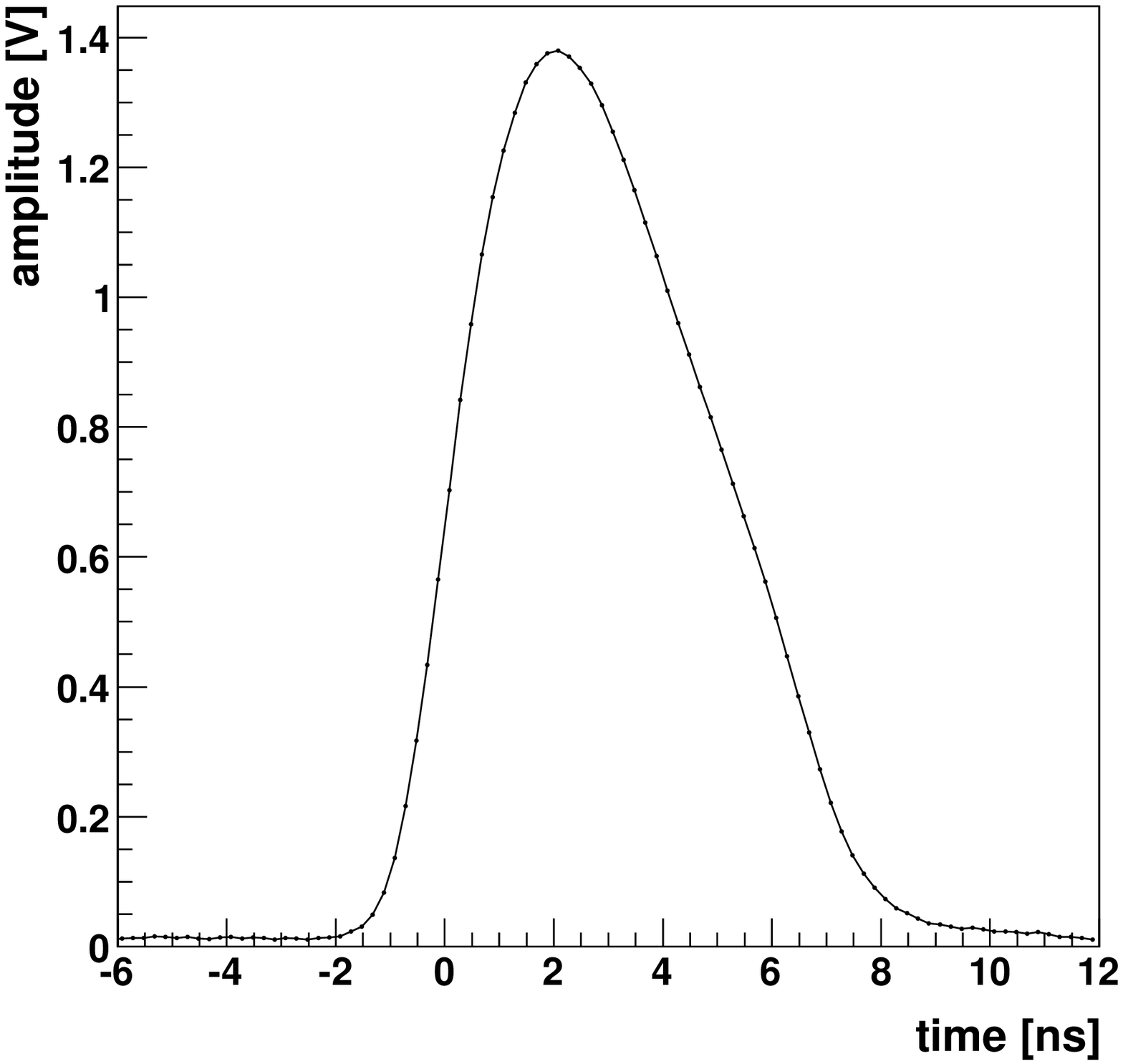}
\end{tabular}
\vspace{-0.1 cm}
\caption{Left: Light amplitude as a function of the azimuth angle for
different cut depths. Right: Timing profile of a single LED pulse.}
\vspace{0.5 cm}
\label{fig:aLED}
\end{center}
\end{figure}

Figure~\ref{fig:aLED} (left) shows the light amplitude as function of the
azimuth angle for different cut depths. The right plot in
figure~\ref{fig:aLED} shows the time distribution of a single LED pulse.

%%%%%%%%%%%%%%%%%%%%%%%%%
%%%%%%%%%%%%%%%%%%%%%%%%%%%

As mentioned in the previous subsection, the amount of energy per pulse
emitted by the LED can be varied by changing the DC voltage from 0 to
24~V. Below 8~V the amount of light emitted is almost negligible. For a DC
level of 24~V the energy per pulse emitted is at least 150~pJ, which
corresponds to the emission of approximately $4
\cdot 10^{8}$ photons.

\subsection{Assembly, synchronisation, testing and integration}
\label{chap: ASSEMBLY}
\begin{figure}[htpb]
\begin{center}
\begin{tabular}{c c}
\hspace{0 cm}
\includegraphics[width=0.5\linewidth, angle=0]{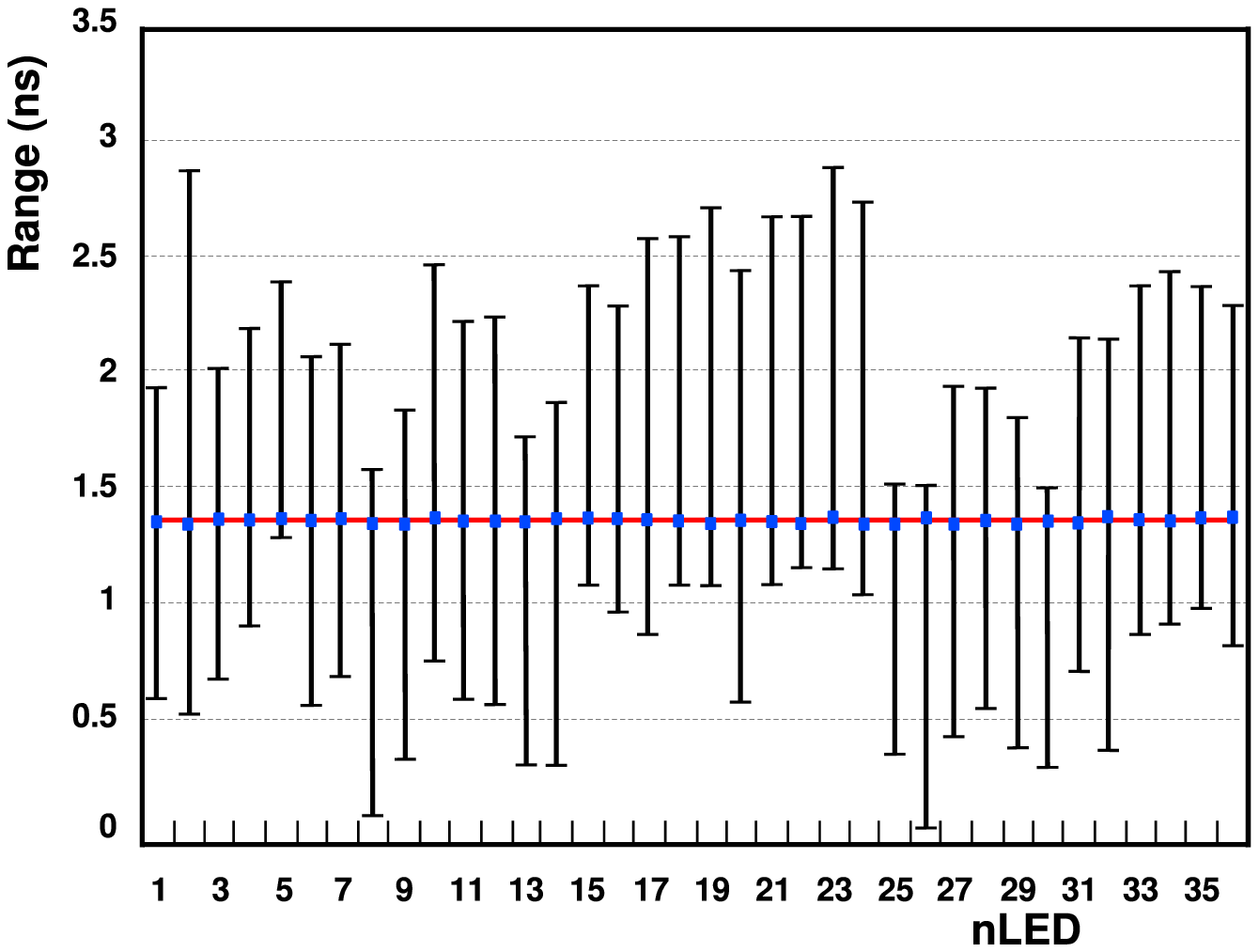}
\includegraphics[width=0.4\linewidth, angle=0]{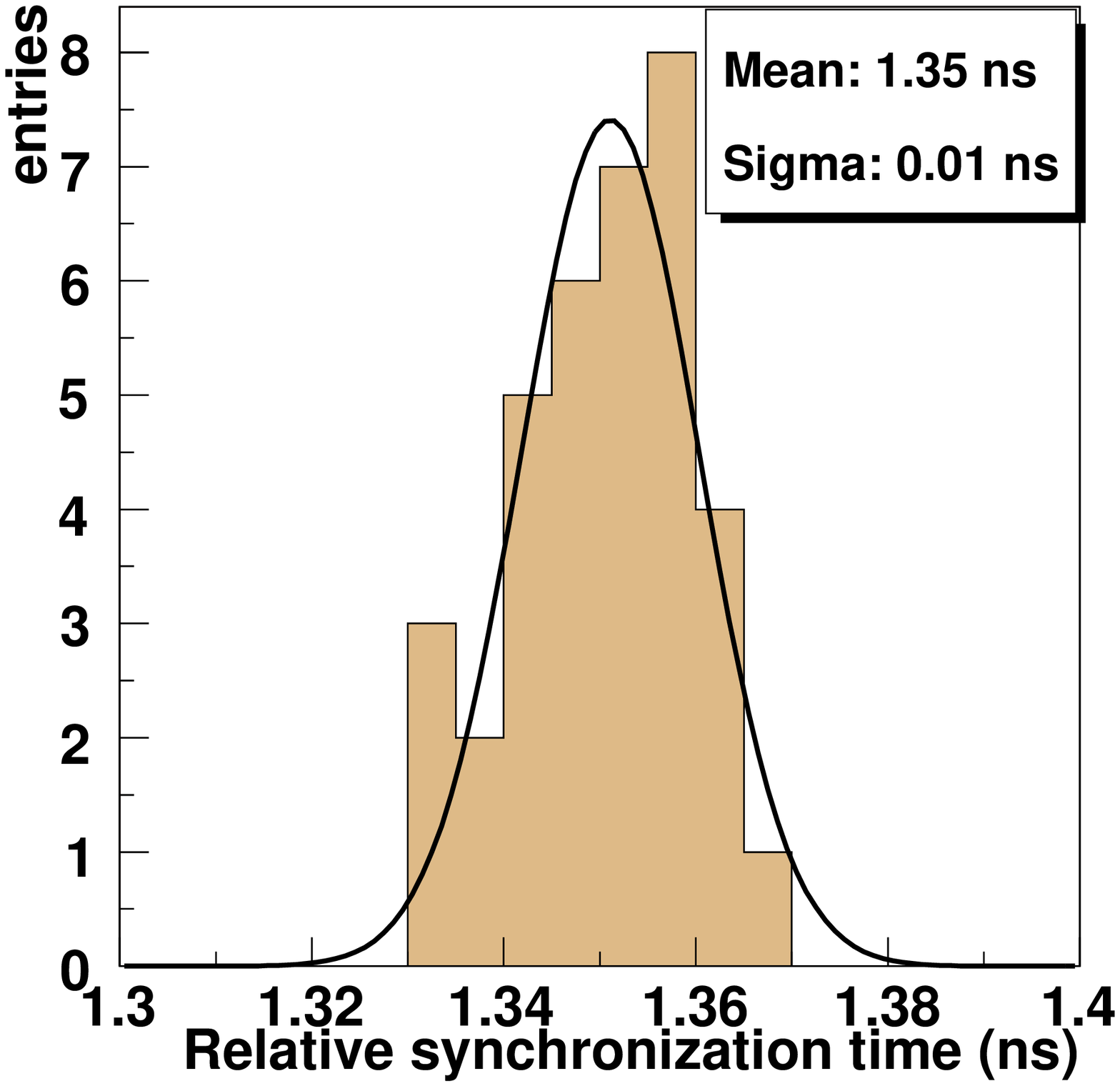}
\end{tabular}
\vspace{-0.1 cm}
\caption{Left: Emission time ranges of the 36 LEDs of 
 a typical LED Beacon. The horizontal straight line indicates the desired
common emission time (reference time), the small squares show the final measured emission
times. Right: Distribution of the final emission times.}
\vspace{0.5 cm}
\label{fig:range}
\end{center}
\end{figure}

Once an LED Beacon is assembled, its 36 LEDs have to be synchronised by tuning
the variable capacitor in each of its pulser circuits. This operation takes
place in three steps. First the achievable time range of light emission is
measured for each LED by adjusting the variable capacitor to its upper and
lower limits. Then, taking advantage of the overlap of these time
ranges, a common reference time for all LEDs is chosen. Finally, the
capacitors are tuned again to reach this reference time. In the left-hand
plot of figure~\ref{fig:range} the different ranges for each LED pulser for a
typical LED Beacon are shown. The full (red) line indicates where the
synchronisation common reference time was set and the squares (blue)
are the final measured times. The right-hand plot of figure~\ref{fig:range}
shows the distribution of the final emission times. The typical standard
deviation of the emission times is in general a few tens of picoseconds.
Following this synchronisation procedure, the beacon undergoes a series of
thermal cycles in a climate chamber lasting 48 hours in order to guarantee
its stability. The emission times are then remeasured and in exceptional
cases some pulsers are re-synchronised.

%%%%%%%%%%%%%%%%%%%%%%%%%
%%%%%%%%%%%%%%%%%%%%%%%%%%%

\begin{figure}[htpb]
\begin{center}
\begin{tabular}{c c}
\hspace{-0.5 cm}
\includegraphics[width=0.7\linewidth, angle=0]{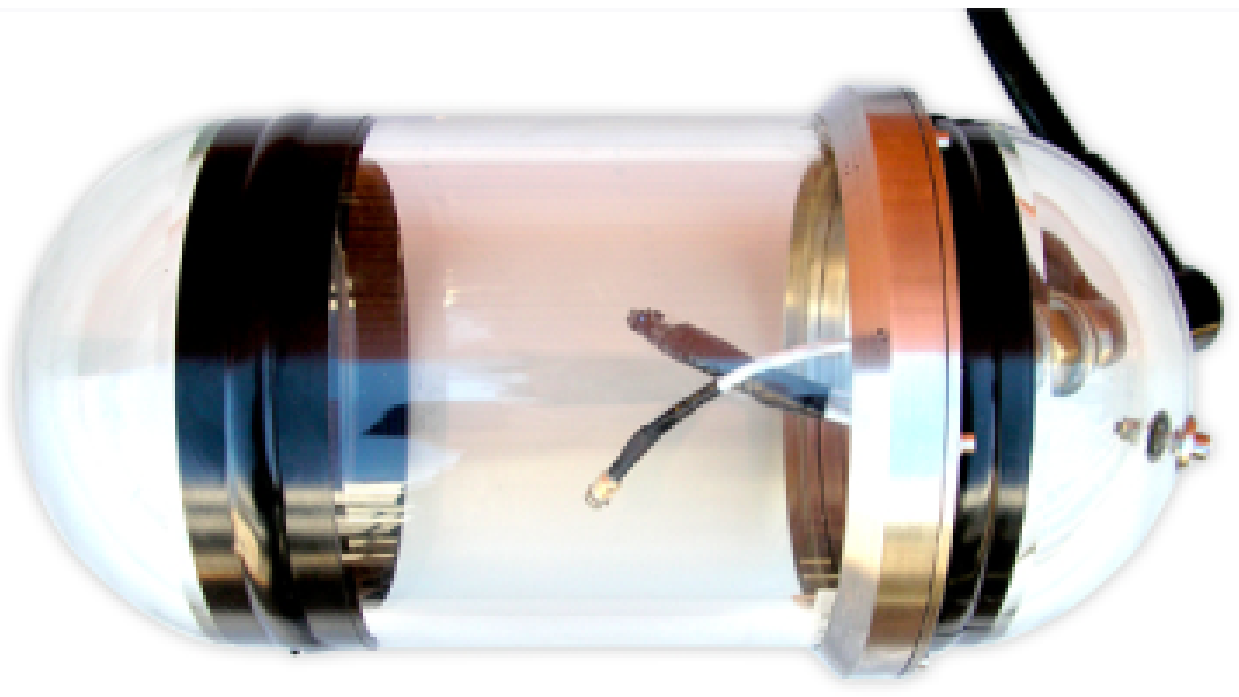}
\hspace{0 cm}
\includegraphics[width=0.3\linewidth, angle=0]{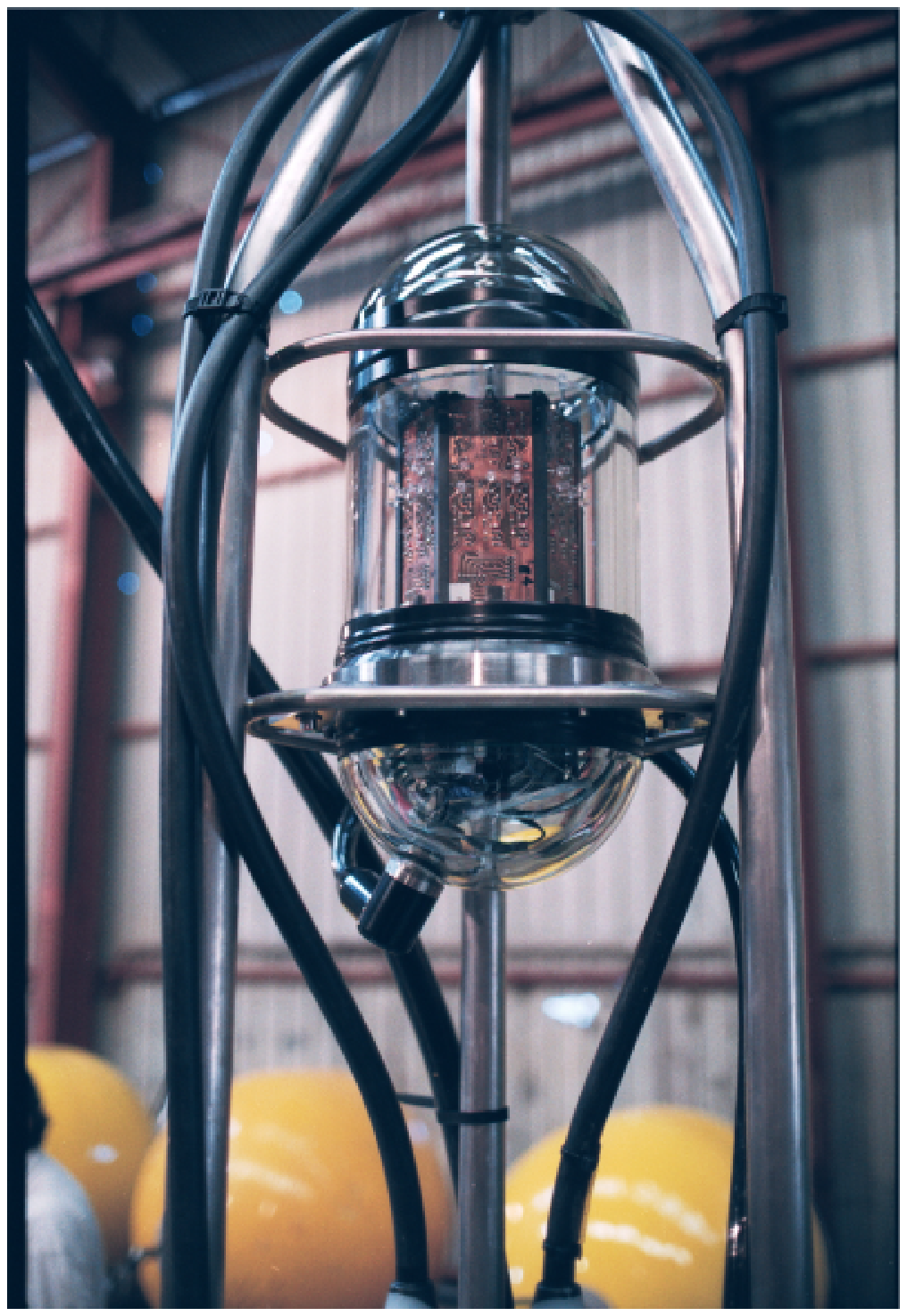}
\end{tabular}
\vspace{-0.1 cm}
\caption{Left: Example of a borosilicate glass container that houses the LED Beacons. 
Right: LED Beacon mounted in the Optical Module Frame.}
\vspace{0.2 cm}
\label{fig: container}
\end{center}
\end{figure}

After synchronisation, the LED Beacon is introduced into the pressure
resistant glass vessel that will house it in the sea.
This is a cylindrical borosilicate glass container commercially
available\footnote{Nautilus Marine Service GmbH, Blumenthalstrasse 15
D-28209 Bremen Germany.}.
It consists of a cylinder and two end-caps 
(figure~\ref{fig: container} left), one of them detachable. All
the parts are in transparent glass except for the titanium flanges that 
hold the two parts together. The overall dimensions of the cylinder
plus endcaps are 210~mm outer
diameter and 443~mm in length. The endcap is supplied with a 22~mm diameter
pre-drilled hole equipped with a penetrator on the outside of the cylinder
and connecting cables on the inside. 
 The LED Beacon is mechanically attached to the detector lines by a
collar mounted on the Optical Module Frame (OMF). It is held
vertically at a specific location above the triplet of OMs 
(see figure~\ref{fig:
container} right) and fixed to the structure combining the 6-fold 
symmetry of the beacon and the 3-fold symmetry of the OMF so as to
minimise shadowing.

As already mentioned, the lines undergo a calibration procedure at the
integration sites using a common laser source. The PMT of the LED Beacon is
calibrated simultaneously with the PMTs in the OMs using a dedicated fibre from this common
source. This calibration facilitates the measurement of their relative
shift in the arrival times.

~\\
%%%%%%%%%%%%%%%%%%%%%%%%%
%%%%%%%%%%%%%%%%%%%%%%%%%%%

\subsection{Configuration, control and monitoring}
\label{chap: LEDCONTROL}
The LED Beacon motherboard controls the operation of the beacon
at the most basic level. 
Its main component is the UNIV1 card, a control board 
used throughout ANTARES which is based on a 17C756 PIC controller. 
Communication with the motherboard takes place on an RS485 serial link protocol MODBUS.

The motherboard main functions are 1) to set the DC level supplied to the
LEDs (from the +48~V input voltage via DC/DC converter) and thus the
intensity of the light pulses emitted, 2) to select the faces or group of
faces that will be flashed, 3) to select any of the three groups of LEDs or
the combination thereof that will flash, 4) to set the PMT gain control
voltage, and 5) to monitor the voltage supplied to the LEDs and the ambient
temperature as measured by a sensor.
  
  The configuration of all the LED Beacons in the detector is performed
through the general RunControl program~\cite{Readout}. Different
configurations corresponding to the set of possible faces and groups of LEDs
that are requested to flash and the required PMT and LED voltages, are stored
in a database. In the configuration stage at the start of a calibration run,
a given configuration (a {\it run set-up}) can then be selected and
downloaded. In addition, a graphical user interface written in Java allows
the expert user to communicate directly with the beacons and request their
status or change their configuration during the run by generating the
necessary Slow Control commands. This latter method is very rarely used
during normal data taking, but is employed in the laboratory for test or
debugging purposes.

\section{The Laser Beacons}
\label{chap: LASERsystem}
\begin{figure}[htpb]
\begin{center}
\begin{tabular}{c}
\hspace{0 cm}
\includegraphics[width=0.6\linewidth, angle=0]{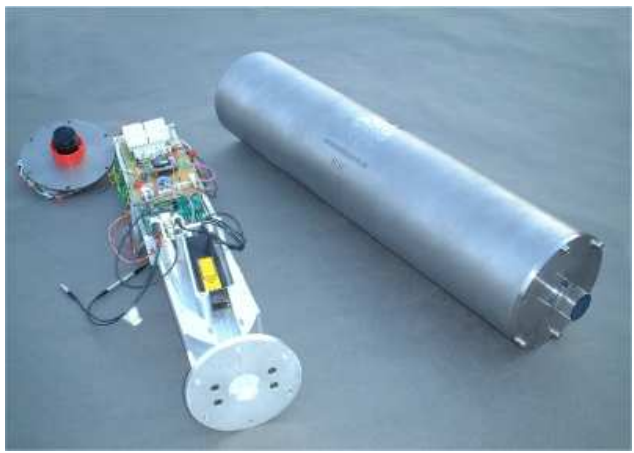}
\end{tabular}
\vspace{-0.1 cm}
\caption{A Laser Beacon dismantled: On the left the inner mechanics
  holding the laser head and its associated electronics; on the right the
  pressure-resistant titanium container that houses the equipment.  On the
  top endplate (right part of the picture) of the container the quartz
  cylinder that prevents sedimentation effects can be seen.}
\vspace{0.5 cm}
\label{fig: laser}
\end{center}
\end{figure}

The Laser Beacons emit high intensity, short duration pulses of light and
will be located at the bottom of a few lines, attached to their Bottom String
Socket (BSS), i.e. the mechanical structure that anchors the line to the sea
bed.  At present, one Laser Beacon has been installed in the so-called MILOM
line~\cite{Milom} (see next section). Figure~\ref{fig: laser} shows a general
view of the Laser Beacon and its components. The Laser Beacon points upwards
so that the emitted light can reach nearby lines.  Since the intensity and
time profile of the light pulse are defined by the intrinsic properties of
the laser, an extensive study of different laser types was made. The selected
model is described in the following subsection.

%%%%%%%%%%%%%%%%%%%%%%%%%
%%%%%%%%%%%%%%%%%%%%%%%%%%%

\subsection{The laser}
\label{chap: LASER}
The main component of the Laser Beacon is a diode pumped Q-switched Nd-YAG
laser which produces short pulses 
with a time duration less than 1~ns (FWHM) and a total
energy of $\sim$1 $\mu$J. 
The laser model selected is 
the NG-10120-120\footnote{Nanolase, presently part of
JDS Uniphase Corp., 430 N. McCarthy
 Blvd.  Milpitas, CA
95035 United States.} which emits at 532~nm after frequency doubling
of the original Nd-YAG wavelength of 1064~nm.

The laser is very compact. Its head dimensions in mm are $144\times37.4\times30$. The laser can be operated in a non-triggered mode
at a fixed frequency (around 15~kHz) or in a triggered mode with a variable
trigger frequency. In the latter mode, which is the one being used in
ANTARES, the laser is triggered when a TTL signal arrives at the device
through a connection in the rear panel of the power supply. Since the laser
is passively Q-switched, the delay between the trigger signal and the light
pulse emission is of the order of microseconds and the pulse to pulse jitter is of the order of a few hundred nanoseconds. The actual
time of laser emission is obtained thanks to a fast photodiode integrated
into the laser head.

Once the laser shot is produced, the built-in photodiode sends back a
signal which is passed to an ARS chip located in the String Control Module
(SCM), the electronics container similar to the LCM located on the BSS. The
current that feeds the pumping diode is switched off and the system
waits for the next trigger signal.

The power supply delivered with the laser was refurbished in order to comply
with the technical requirements of the experiment and, at the same time, to
accommodate the whole apparatus into a smaller space. The signal from the
photodiode is reshaped electronically to fulfill the constraints imposed by
the front-end electronics of the experiment.

\begin{figure}[htpb]
\begin{center}
\begin{tabular}{c c}
\includegraphics[width=0.5\linewidth, angle=0]{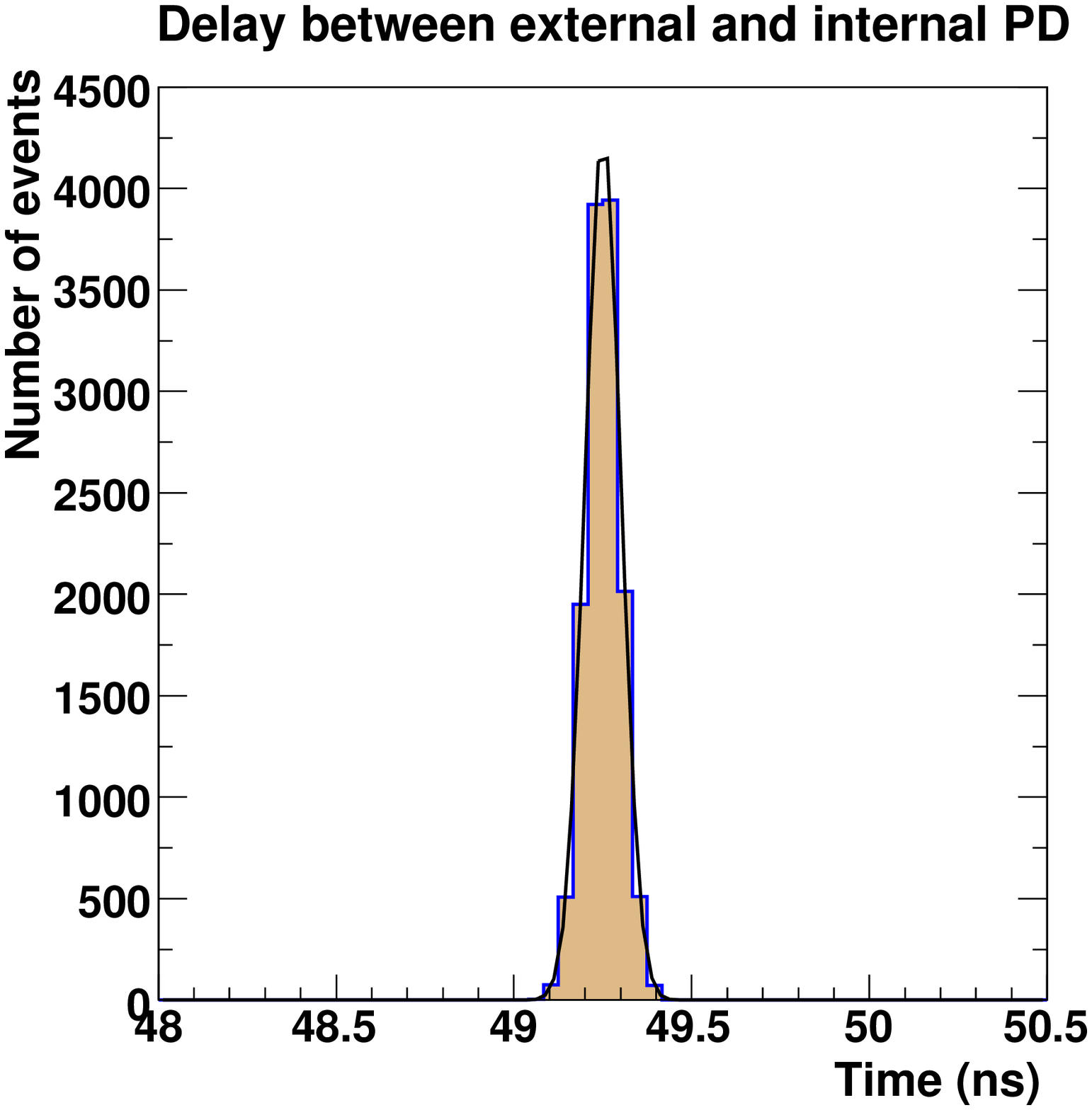}
\includegraphics[width=0.5\linewidth, angle=0]{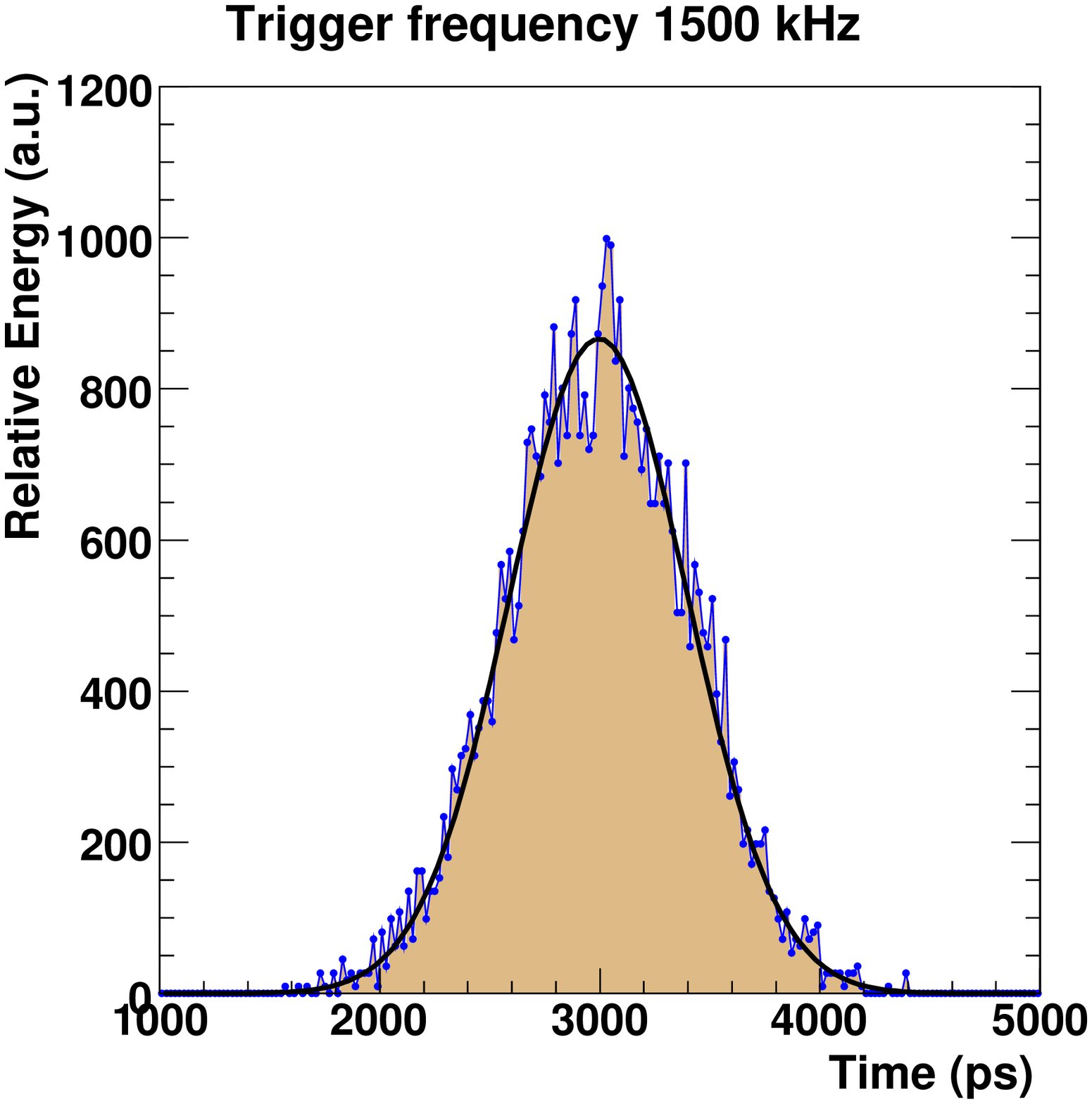}
\end{tabular}
\begin{tabular}{c c}
\includegraphics[width=0.5\linewidth, angle=0]{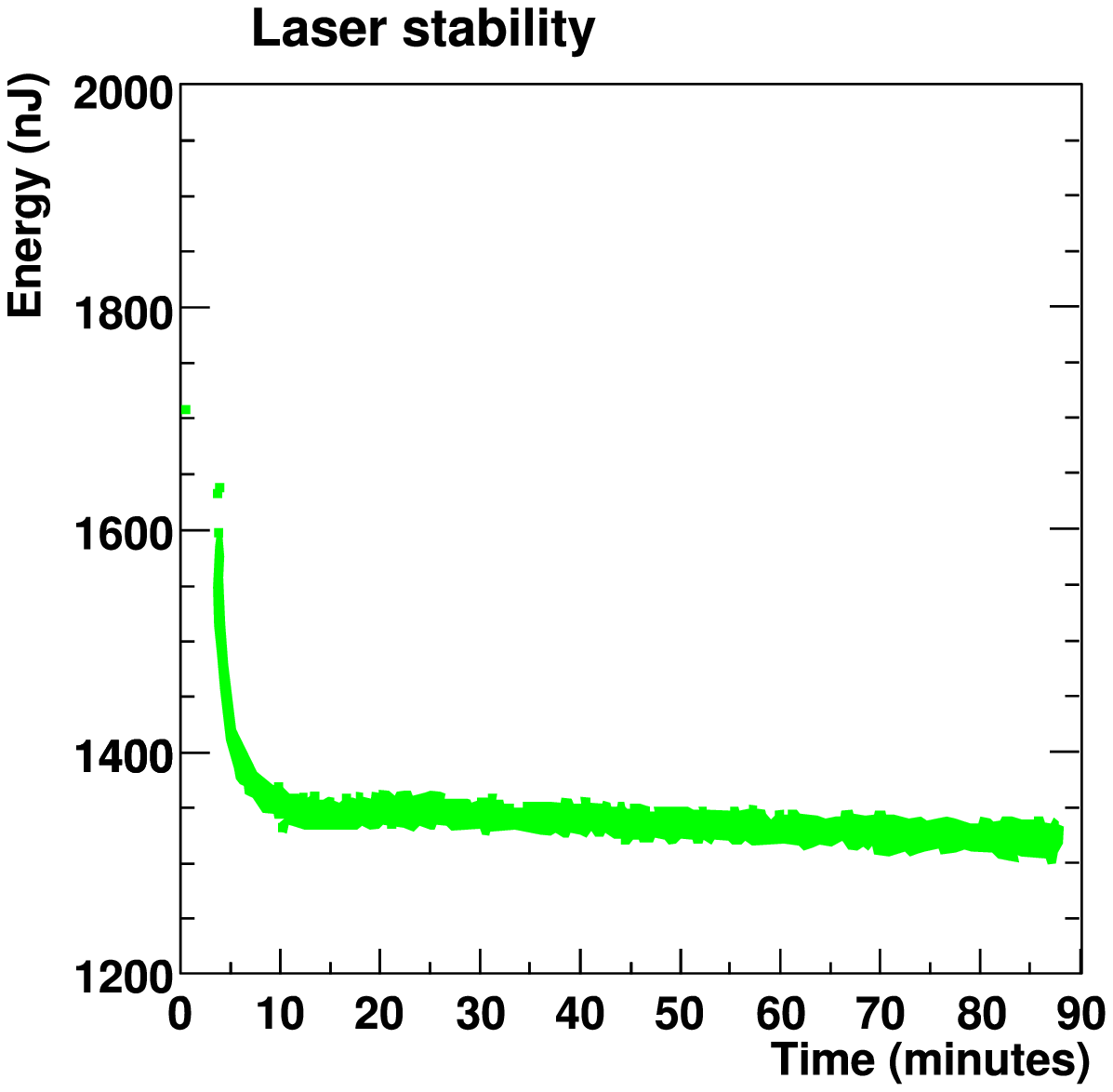}
\includegraphics[width=0.5\linewidth, angle=0]{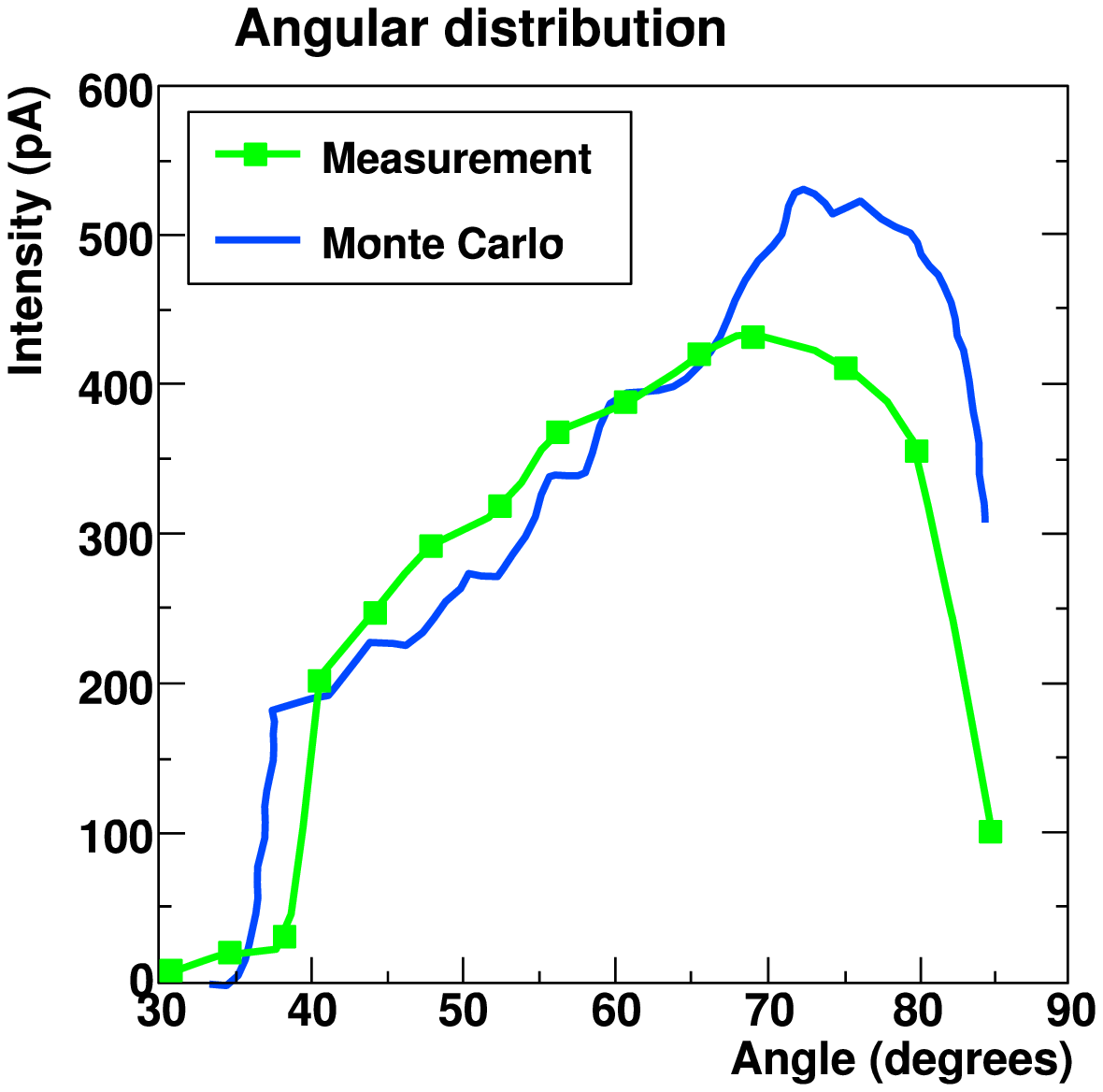}
\end{tabular}
\vspace{0 cm}

\caption{Top-left: Distribution of the difference of the time of
  emission of the laser pulse as measured by an external Newport 818-BB-20
photodiode (PD) and the laser built-in photodiode. Top-right: Timing profile of
the laser pulse as measured by a Hamamatsu streak camera. Bottom-left:
Energy of the laser as a function of time after switch-on. Bottom-right: Angular distribution of
the Laser Beacon in water and comparison with Monte Carlo predictions (see section~\ref{chap: LASERB2}).}
\label{fig: shape}
\end{center}
\end{figure}

In order to characterise the relevant features of the Nd-YAG
laser, a thorough study of the main laser parameters was made:

\begin{enumerate}

\item 
  The intrinsic jitter of the Q-switching mechanism gives rise to a jitter in
  the laser pulse emission time of a few hundred nanoseconds. It was,
  therefore, necessary to confirm that the time recorded by the internal
  photodiode was sufficiently accurate for our needs.  Several fast external
  photodiodes (Newport 818-BB-20, Alphalas UPD-200-SP, Hamamatsu S5973-01)
  were employed to estimate the accuracy in the emission time given by the
  internal photodiode. Figure~\ref{fig: shape} (top-left) illustrates the
  difference in emission time as measured by an external Newport 818-BB-20
  photodiode and the internal built-in photodiode.  The standard deviation of
  the distribution is 50~ps (the position of the peak is immaterial, it
  depends on delays that will be determined by calibration).

\item The pulse shape was measured using a Hamamatsu streak camera. In
figure~\ref{fig: shape} (top-right) an example of a pulse of the laser as
sampled by the streak camera (5~ps resolution) is shown. The FWHM of the
pulse is determined to be smaller than 0.8~ns. The time shape profile is
smooth and not far from Gaussian. As expected, the timing features of the pulse did
not change with the three trigger frequencies studied, namely 100~Hz, 1.5~kHz
and 10~kHz.

\item Different energy measurements were also performed 
with a special device\footnote{A photodiode head Model PD10 and a laser power
  meter LaserStar from Ophir, Laser Measurement Group Ophir Optronics
  Inc. 260-A Fordham Road Wilmington, MA 01887 United States.} capable of
  measuring the energy of each pulse.  As can be seen in figure~\ref{fig:
  shape} (bottom-left), after a warm-up period of $\sim$5~minutes, the energy
  yield per pulse (close to 1.3 $\mu$J) becomes stable ($<\pm 3$\%).

\end{enumerate}

%%%%%%%%%%%%%%%%%%%%%%%%%%%
%%%%%%%%%%%%%%%%%%%%%%%%%%%

\subsection{The Laser Beacon mounting}
\label{chap: LASERB2}

The laser is housed in a cylindrical titanium container 705~mm in length and
170~mm in diameter (see figure~\ref{fig: laser}).  The bottom endcap holds
the penetrator of the cable connectors. Inside the container, an aluminium
inner frame holds the laser and its associated electronics. The laser beam
points upwards and leaves the container through an opening in the top
end-cap. In this opening there is an optical diffuser comprising a flat disk
diffuser\footnote{ORIEL model 48010, Newport Corporation-Oriel 150 Long Beach
lvd. Stratford, CT 06615 United States.}  with a thickness of 2.2~mm and a
diameter of 25~mm that spreads light beam out following a cosine distribution, so
that the light can reach the surrounding lines.

\begin{figure}[htpb]
  \begin{center}
    \begin{tabular}{c c}
      \includegraphics[width=0.5\linewidth, angle=0]{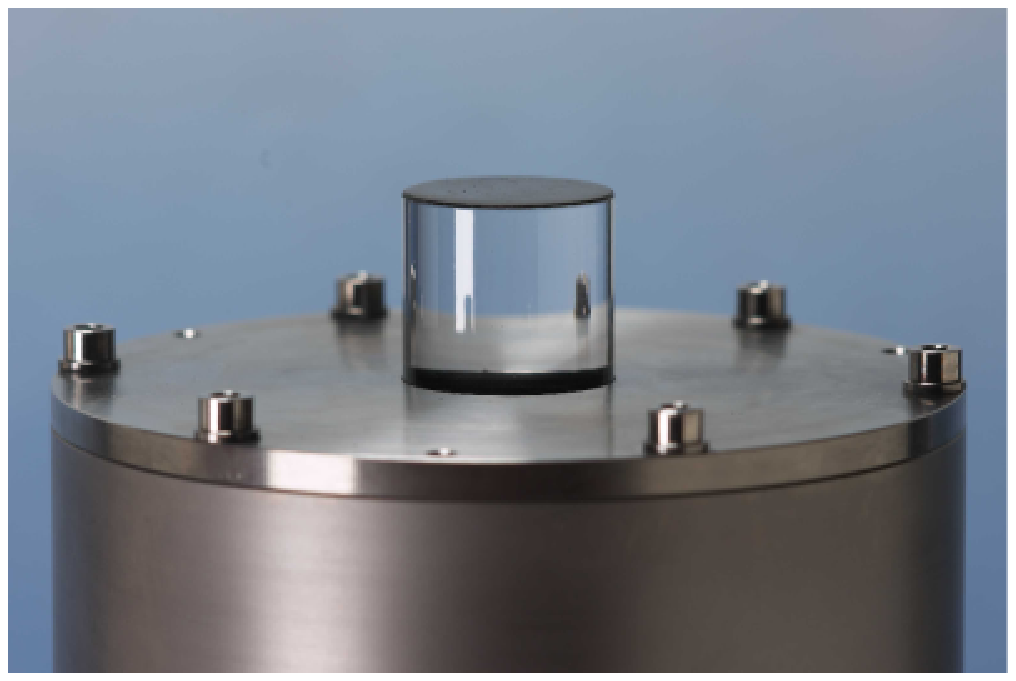}
      \hspace{0 cm}
      \includegraphics[width=0.5\linewidth, angle=0]{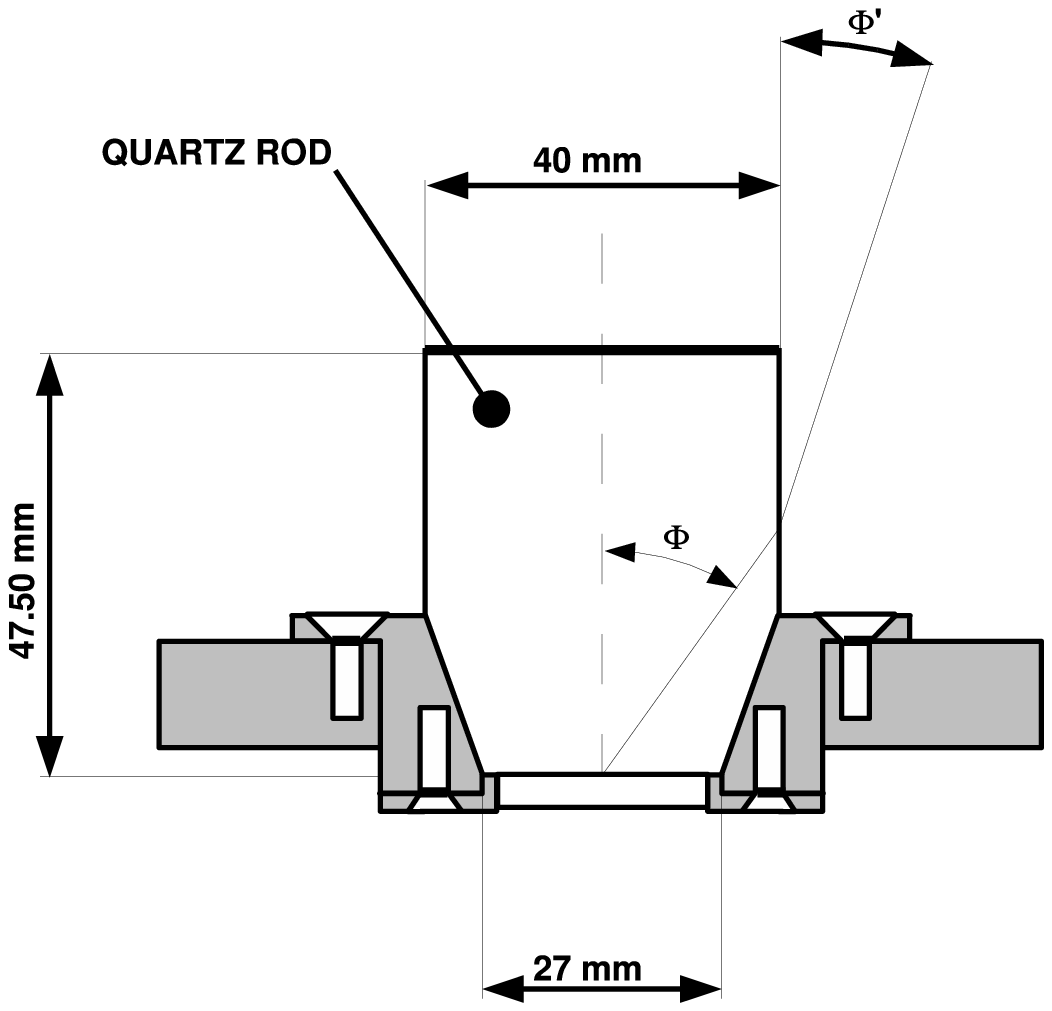}
    \end{tabular}
    \vspace{-0.1 cm}
    \caption{Left: Upper endplate of the Laser Beacon container
     with the quartz cylinder used to avoid losses due to
      biofouling. Right: Schematic path of the outgoing light.}
    \label{fig: rod}
  \end{center}
\end{figure}

In order to minimize transmission losses due to underwater sedimentation and biofouling\footnote{the accumulation of
micro-organisms, mostly bacteria, on outer surfaces and fallen sediment subsequently
adhering to surfaces~\cite{SED}.} (the laser is more affected by this effect
since it is pointing upwards), a quartz cylinder was bonded to the upper
surface of the diffuser as can be seen in figure~\ref{fig: rod} (left). The
upper surface of this cylinder is coated with a black, water resistant epoxy
layer. The light then leaves the cylinder through the vertical wall where
biofouling is negligible (see scheme in figure~\ref{fig: rod} (right)). Due to
Snell's law, the cosine distribution is conserved when the light leaves the
cylinder through its vertical wall.

The dimensions of the cylinder were chosen to be 40~mm in diameter and 47~mm
in length. These dimensions, which together with the refractive index of
quartz ($n = 1.54$) and water ($n = 1.34$) determine the maximum and minimum angle
of the outgoing light, were selected to maximise the number of storeys
illuminated in the closest lines, while taking into account the technical
contraints of the cylinder fixing due to the high pressure.

  The distribution of the outgoing light was measured by immersing the Laser
 Beacon into a large water tank that had a PMT on its upper cover and
 inclining the beacon at different angles with respect to the vertical. In
 figure~\ref{fig: shape} (bottom right) the resulting angular distribution is
 shown together with the result of a simulation. As can be seen, they are
 in quite good agreement except for high angles, where there are edge effects due
 to a bevel in the quartz cylinder not reproduced by the Monte Carlo.

%%%%%%%%%%%%%%%%%%%%%%%%%
%%%%%%%%%%%%%%%%%%%%%%%%%%%

 The laser diode is fed with 5~V and 750~mA supply. As in the case of the
 LED Beacon, a UNIV1 micro-controller board is used to provide the
 correct power supply parameters, execute the corresponding switch on and off
 signals and monitor the measurements given by temperature and
 humidity sensors inside the container. A thermo-electrical cooling
 system based on Peltier cells ensures the correct operating
 temperature of the laser head.

\section{First results of the Optical Beacon system}
\label{chap: results}
An instrumentation line with Optical Modules (MILOM)~\cite{Milom} and the
first two complete lines of the detector, {\it Line 1} and {\it Line 2}, were
already in operation at the ANTARES site at the end of 2006. The three
lines are connected to a Junction Box which in turn is linked to the shore by
an electro-optical cable. The lines are operated from a control room situated
at La Seyne-sur-Mer, near Toulon (France), and data are regularly taken since their connection.

The MILOM has one operating LED Beacon at storey 1 and one Laser Beacon
located at the bottom of the line. {\it Lines 1} and {\it 2} are equipped
with four LED Beacons at storeys 2, 9, 15 and 21 (numbered from bottom to
top). Dedicated data taking runs in which one or several of the Optical
Beacons are flashed are regulary performed. Only data from the Optical Beacon
system in {\it Line 1} and MILOM are presented here. 

Light emitted from the various Optical Beacons is detected by the OMs on the two
lines. The distribution of the arrival times of the light at the OMs is
presented below. Further details of this analysis will be the subject of a
subsequent paper.

%%%%%%%%%%%%%%%%%%%%%%%%%
\begin{figure}[htpb]
\begin{center}
\begin{tabular}{c}
\hspace{-1 cm}
\includegraphics[width=1.1\linewidth, angle=0]{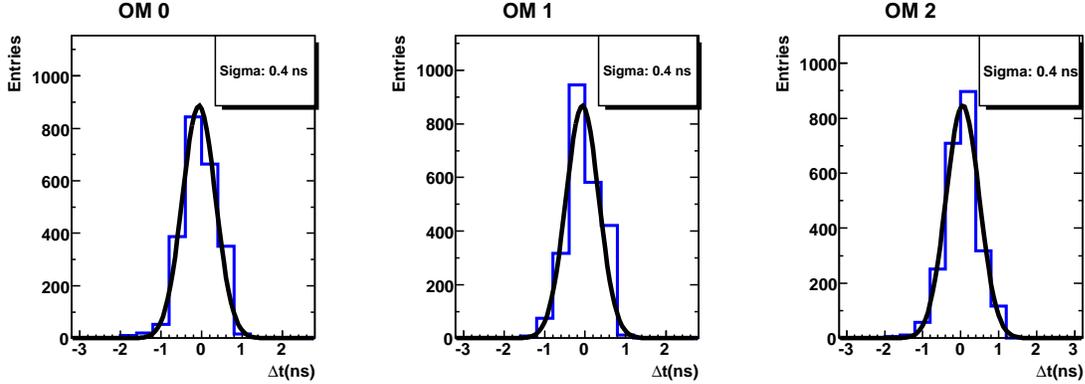}
\end{tabular}
\vspace{-0.1 cm}
\caption{Distributions of the difference between the time of
arrival of the light at the three OMs of storey 3 of {\it Line 1} 
and the time of emission from the LED Beacon in storey 2 on the 
same line.}
\label{fig: Line1}
\end{center}
\end{figure}
%%%%%%%%%%%%%%%%%%%%%%%%%%%

 Figure~\ref{fig: Line1} shows the distribution of the difference between the
time of the signal recorded by each of the three OMs on the third storey of
{\it Line 1} and the time of emission of the pulse at the LED Beacon in the
storey below as recorded by the beacon's PMT. Together with the time
distributions, a Gaussian fit is shown. The values of the Gaussian width are
also indicated. The means have been set to zero.

Due to the short distance between the beacon and the OMs ($\sim 13.4$~m) and
to the high intensity at which the beacon was operated, contributions to the
width of the time distributions from the transit time spread in the PMT, from
the signal pulse walk, from light scattering in the water and from line
movements are all negligible. The width of the time distribution, which is
$\sim 0.4$~ns, is therefore a measure of the resolution of the read-out
electronics which is compatible with the time resolution measured
in reference~\cite{ARS}.

%%%%%%%%%%%%%%%%%%%%%%%%%
\begin{figure}[htpb]
\begin{center}
\begin{tabular}{c}
\hspace{-1 cm}
\includegraphics[width=1.1\linewidth, angle=0]{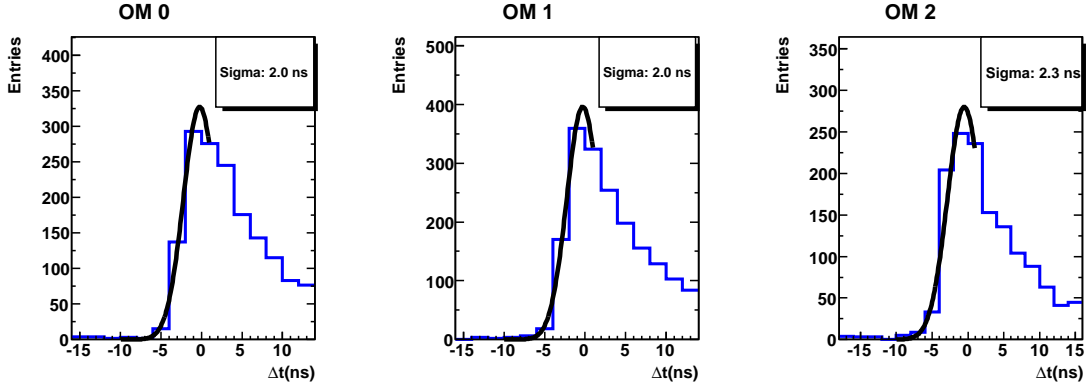}
\end{tabular}
\vspace{-0.1 cm}
\caption {Distributions of the difference between the arrival time
of light at the three OMs of storey 2 of the MILOM 
and the emission time from the Laser Beacon at the bottom 
of the same line.}
\label{fig: MILOM}
\end{center}
\end{figure}
%%%%%%%%%%%%%%%%%%%%%%%%%%%

Figure~\ref{fig: MILOM} shows the distributions of the difference between
the detection time of the signal at the three OMs of the second storey
of the MILOM and the emission time of the pulse from the Laser Beacon
as provided by the built-in photodiode. In this case, 
due to the distance between the Laser Beacon and the OMs ($\sim109$~m)
a long tail due to scattering is clearly visible. The sigmas of a Gaussian
fit to the rising edge of the distributions are shown. The rise-time of the
distributions is the outcome of the convolution of the rise-time 
of the laser pulse itself and the effect of scattering on the early photons. 
No correction has been applied to the data. The position of the peaks of the distributions
have been set to zero.

\begin{figure}[htpb]
\begin{center}
\begin{tabular}{c}
\hspace{-1 cm}
\includegraphics[width=1.1\linewidth, angle=0]{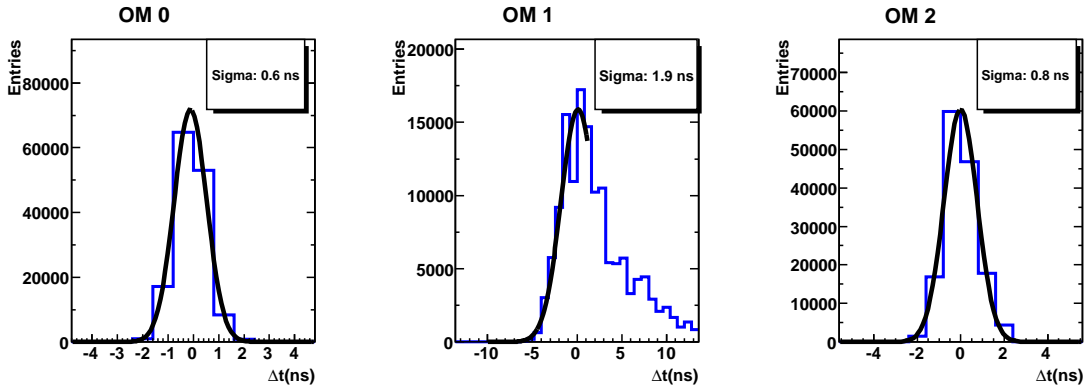}
\end{tabular}
\vspace{-0.1 cm}
\caption {Distributions of the difference between the arrival time of
the light at the three OMs in {\it Line 1} storey 3 and the emission time
from the LED Beacon in the first storey of the MILOM.}
\label{fig: MILOM+Line1}
\end{center}
\end{figure}

The Optical Beacon system is also used for time calibration between different
lines. Figure~\ref{fig: MILOM+Line1} shows the distribution of the difference
between the detection time of the signal at the OMs in {\it Line 1} third
floor and the emission time of the light by the LED Beacon located in the
MILOM first storey. The distance between the triplet and the Optical Beacon
is about 80~m. Due to the orientation of the triplet in the storey, one of
the OMs looks in the opposite direction to the MILOM and therefore can only
detect scattered photons which explains the tail in the time
distribution. The results of the Gaussian fits to the full distributions, or
its rising edge when scattering is present, are shown.

%%%%%%%%%%%%%%%%%%%%%%%%%
%%%%%%%%%%%%%%%%%%%%%%%%%%%

\section{Summary and conclusions}
\label{chap: summary}

 In order to ensure the high angular resolution that is aimed by ANTARES, an exhaustive R\&D
study to develop different time calibration systems has been carried
out. These systems must guarantee a relative time resolution at the
sub-nanosecond level between the Optical Modules of the detector. In this
article the main features of the Optical Beacon calibration system of the
ANTARES detector have been described. This system uses short light pulses
produced by LEDs or lasers to measure time differences and therefore
calibrate the timing of the detector. The design and construction of these
Optical Beacons together with the tests performed in the laboratory have been
described. The first results of some of these devices located at 2500~m depth
have been given. With these calibration systems it has been demonstrated
that, in situ, ANTARES achieves a sub-nanosecond time resolution.

\renewcommand{\thesection}{}
\section{Acknowledgements}
\label{}

The authors acknowledge the financial support of the funding agencies: Centre
National de la Recherche Scientifique (CNRS), Commissariat \`a l'Energie
Atomique (CEA), Commission Europ\'eenne (FEDER fund and Marie Curie Program), R\'egion Alsace
(contrat CPER), R\'egion Provence-Alpes-C\^ote d'Azur, D\'epartement du Var
and Ville de La Seyne-sur-Mer, in France; Bundesministerium f\"ur Bildung und
Forschung (BMBF), in Germany; Istituto Nazionale di Fisica Nucleare (INFN),
in Italy; Stichting voor Fundamental Onderzoek der Materie (FOM), Nederlandse
organisatie voor Wetenschappelijk Onderzoek (NWO), in The Netherlands;
Russian Foundation for Basic Research (RFBR), in Russia; Particle Physics and
Astronomy Research Council (PPARC), in U.K.; National Authority for
Scientific Research (ANCS) in Romania; Ministerio de Educaci\'on y
Ciencia (MEC), in Spain.

% \appendix

% \section{}
% \label{}

\end{document}